# A hypothesis in evolutionary biology

Vincenzo De Florio, 2018-02-11

**Abstract**: The classic Trivers-Willard hypothesis suggested the existence of means or conditions able to influence or control the sex of the offspring. Here I propose that mechanisms for the alteration of the gender of the offspring could possibly be formulated in terms of a distributed system of messages expressing a change in the environmental conditions. Such messages would provide the biological organization with global and local assessments of the benefits associated with the reproductive investments associated with either genres of the offspring.

Many authors, starting with the classic work by Trivers and Willard [1], hypothesized the existence of "conditions" able to influence or control the sex of the offspring. Ever since the publication of that classic work, researchers have been producing an impressive amount of results that either bring evidence or contradict that hypothesis. This production has been chaotic to say the least, with scholars deriving their conclusions from "facts" ranging from dubious interpretations of microscope images to extremely serious mathematical and statistical models of the many organs and processes at play. I am merely an information scientist, lacking too many important pieces in this trans-disciplinary problem. But I have read several papers on this, and my focus on information led me to a simple observation: in information science, what is really important is *the variation of the signal* rather than the signal itself; in other words, it is a signal's variation that carries information -- it is a message -- while a steady signal it is not. As an example, a diet is a signal while a change in a diet, especially when it is a significant change – it is a message. Evidence corroborating this idea may be found in several works. Professor Elissa Cameron and her team, for instance, found out [2] that, rather than glucose, it is the glucose *gradient* that might play a role in adjusting the offspring's sex ratio in mice. This gradient represents a "message" from the environment, declaring the onset of a more favorable condition. The existence of "messages" of the opposite sign and outcome is also discussed by Professor Kristen Navara, who hypothesizes "a bias toward females during times of stress", mediated by glucocorticoids in her very interesting paper [3]. A stress / glucocorticoids-gradient signals in this case a time where an investment in a male offspring would not be advisable.

Testosterone gradients are another example. It is well known how such changes represent "boosts" that can lead to more daring behaviors. The late Dr. Valerie J Grant, in several articles (e.g. [4]) and her book ([5]) suggests that higher levels of testosterone might be associated with a higher sex ratio. Results are a little contradictory though, and I suspect that this might be due to measurements of the signal and not of its variations. Also it is not clear to me whether testosterone would have a causal role or rise "simply" as a side effect of some "first cause."

Other examples of such gradients may be found in the huge bibliography of Dr. William H. James, who drew several hypotheses of endocrine mechanisms to control or influence the sex ratio (see e.g. [6], [7]).

Of course, if we assume that such messages are actually "there," a logical question would be: *Which type of messages* are at play here? My answer comes from considering the theory of a fractal organization of the all – as one can find, for instance, the human body. The latter is composed of a primary unit "personifying" the whole – the brain – and a sophisticated hierarchy of organized systems, each of which is further composed of hierarchies of organized sub-systems. The overall functioning of the human body requires messages flowing from one end to the other, signaling actions that need to be taken when certain conditions are established; hormonal messages are a typical example of this large distributed organization.

My conclusion is then that mechanisms for the alteration of the gender of the offspring could possibly be formulated in terms of a distributed system of said messages. Perhaps a distributed system of **fractally organized** messages! Messages that is with a scope that ranges from the micro to the macro, signaling waves of actions that could "tune" our internal mechanisms after the external conditions expressed by the environment. In a number of my papers (e.g. [7], [8], [9]), I have suggested that said mechanism could be modeled, in sociotechnical systems, in terms of evolutionary game theory -- which I believe is very much in line with what had been theorized by Trivers and Willard. In fact, my conjecture is that particular messages from the "whole" (the brain) to the "parts" (the systems of organs) are meant to provide contextual information to the latter ones in order to adjust their local action in such a way as to favor a global condition matching the passed information. Local conditions favoring the birth of an offspring with an optimal evolutionary advantage are maybe the outcome of those messages. As a response, "thinner" messages would request the onset of other conditions of a "lesser" (hierarchical) scope. In this view, one could say that Dr. James' hypotheses characterize the endocrine nature of a "layer" of those messages.

As Dr. James himself expressed multiple times  [10], the verification of an hypothesis such as mine would require the collaboration of many a specialist in domains I'm not acquainted with. I hope this short text may produce some interest and trigger exchange of ideas and chances for collaboration.